\documentclass[journal]{IEEEtran}
\usepackage{overpic}
\usepackage{graphicx}
\usepackage{cite}
\usepackage[numbers,sort&compress]{natbib}
\usepackage{booktabs}            

\usepackage{algorithmic}
\usepackage{natbib}
\usepackage[justification=centering]{caption}

\usepackage{url}
\usepackage{stfloats}

\usepackage{rotating}

\usepackage{cite}
\usepackage[colorlinks,linkcolor=red,anchorcolor=red,citecolor=blue]{hyperref}    

\usepackage{hyperref}

\usepackage{algorithm}

\title{Intelligent Processing in Vehicular Ad hoc Networks: a Survey}
\author{Yang~Liu
\thanks{Yang Liu is with the School of Information Engineering, Guangdong University of Technology, Guangzhou,
Guangdong, 510006 China e-mail: (liuyang@gdut.edu.cn).}}

\graphicspath{{figures/}}

\begin{document}
\bibliographystyle{unsrt}
\maketitle

\begin{abstract}
The intelligent Processing technique is more and more attractive to researchers due to its ability to deal with key problems in Vehicular Ad hoc networks. However, several problems in applying intelligent processing technilogies in VANETs remain open. The existing applications are comprehensively reviewed and discussed, and  classified into different categories in this paper. Their strategies, advantages/disadvantages, and performances are
elaborated. By generalizing different tactics in various applications related to different scenarios of VANETs and evaluating their performances, several promising directions for future research have been suggested.
\end{abstract}

\begin{IEEEkeywords}
 Vehicular Ad-hoc Network (VANET); Intelligent Processing; Machine Learning; Fuzzy Logic; Neural Networks; Data Mining; Game Theory
\end{IEEEkeywords}

\section{Introduction}

Vehicular Ad-hoc Networks (VANETs) is a special use case of wireless Ad hoc networks, focusing on  Vehicle-to-vehicle(V2V) data exchange. In 2001, \cite{toh2001ad} firstly proposed VANETs concept , which defined the technology as "car-to-car ad-hoc mobile communication and networking" applications. Recently, VANETs technology emerges as a hot topic. In the near future , VANETs will significantly improve road safety conditions, increase the whole capacity and lift efficiency of the road transport system. By VANETs, these goals can be accomplished while cutting the cost for both vehicle drivers and the administration.

\begin{figure}[htbp]
\centering
\includegraphics[width=0.5\textwidth]{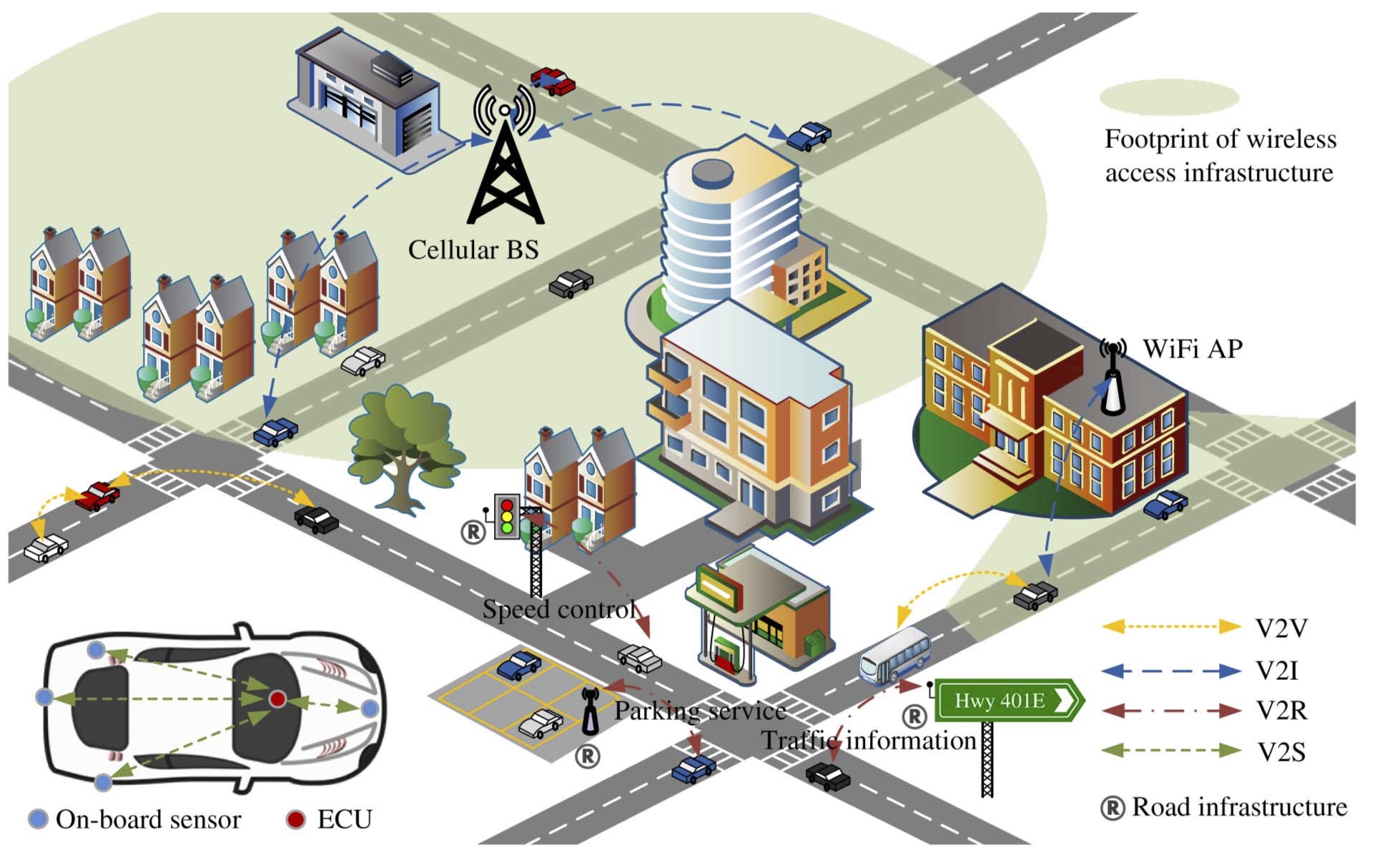}
\vspace{-0.2cm}
\caption{Vehicular Ad Hoc Networks (VANETs)\cite{vanet20181}}
\label{fig_vanet}
\end{figure}

A conceptual view of VANETs is shown in Figure \ref{fig_vanet}, consisting  following entities:
\begin{itemize}
  \item Vehicle. Wireless communication between vehicles
(V2V) and between vehicles and infrastructure access point (V2I) are addressed by VANETs. The transmitted data is mainly generated by vehicles equipping different type of sensors.
  \item Access Point. Fixed and connect to the backbone structure. Vehicles can communicate with the access points by single-hop or multi-hop routing.
  \item Backbone Network Structure. Generally speaking, VANETs can use telecom network as backbone. But the operators often need to build special backbone network in a local area.
\end{itemize}

VANETs have some appealing and attractive features  differ from Ad hoc Networks in general case. They would include, for instance:
\begin{itemize}
  \item Highly dynamic network topology. Vehicle's high speed and high mobility leads to a various issues including the fast changing network topology, density and unbounded size.
  \item Time and integrity Critical. For safety reason, it is extremely important that a correct and up-to-date delivery of information is guarded in VANETs. Actions can be performed accordingly only when information is available as is required.
  \item No power constraints, but budget-conscious. Unlike Wireless Sensor Networks, power supply in vehicle is not a challenge. But if the On Board Unit (OBU) is too complex and expensive, then the attraction of the low-margin auto industry, where scale is key to profitability, is less obvious.
\end{itemize}

Network performance metrics of VANETs includes capacity, QoS insurance, and the redundancies for strict safety constraints. Network capacity is the amount of traffic that a network can handle at any given time. It is very meaningful to some multimedia related V2V applications. Quality of service (QoS) is defined as a set of service requirements, like latency, error rate , priorities for specific types of data, and uptime, that must be met while transporting a packet stream from a source to its destination. For transportation system, the one thing people demand above all others is safety. Application in VANETs must meet tight safety requirements. Failure to comply with traffic safety code can result in civil or even criminal penalties.

Problems come up at once: excessive and contradictory requests from above performance index.  The high-speed nodes (vehicle), fast fading channels, more Doppler shift and rapid change of network topology, together will bring many transmission and networking issues. Compared with traditional wireless networks, the protocols of VANETs will need to modify in a new manner.

Intelligent Processing Technology (IPT) can imitate human intelligence in a way (being human-like rather than becoming human). Compared with "pure" mathematical methods, for example, convex optimization approach\cite{Ghorai2018,Zhou2018,Ma2014,Yu2016,Jeong2018,Zhao2015,Hua2018,Qian2018,Sun2018,Ahnn2014,Liu2014a,Wang2017a,Wang2018a} ,IPT tends to offer faster processing speed while maintaining a reasonable accuracy. Hardware requirement for OBU is simple and inexpensive using IPT. This feature is particularly suited to VANETs. Various studies have established the fact that IPT can play a significant role in VANETs, thus, there is an urgent need to comprehensively study the existing IPT usage cases in VANETs, and summarize their key issues.

This review paper makes three significant contributions to
the field of IPT used in VANET :

\begin{itemize}
\item The main applications of IPT in VANETs are described, with IPT being considered separately to application-specific algorithms.

\item The methods by which the major contemporary and historical algorithms approach  are discussed in detail. Comparative study in each sub-category is also presented.

\item A technical roadmap is presented to illustrate in detail that how new IPT can be successfully deployed in real products.
\end{itemize}

The rest of the paper is organized as follows. Section \ref{sec:B} presents the network model, and IPT classification. Here we give the Road-Map for integrating new IPT to the real VANETs systems. Section \ref{sec:NeMl} introduces the basic ideas and technologies about machine learning techniques involved in VANETs from recent year's publications. In section \ref{sec:flf}, we discuss, classify and compare usage of fuzzy logic techniques in VANETs. Bionic Optimization, Data Mining and Game Theory technologies and their applications are discussed briefly in Section \ref{sec:Bot}. Finally, Section \ref{sec:CFD} concludes the review.

\section{IPT Classification and the Road-MAP for Realization}
\label{sec:B}
Vehicular ad-hoc network applications range from road
safety applications oriented to the vehicle or to the driver, to
entertainment and commercial applications for passengers. We have divided the applications into three major categories:
\begin{itemize}
\item \emph{QoS Guarantee} Key QoS index in VANETs include but not limited to network capacity, end-to-end delay and packet delivery ratio. They are critical for VANETs applications, especially in some specific situation such as multimedia or emergency message dissemination. The most common scenario concerning QoS is routing. There are a lot of IPT using cases in routing for VANETs now.
\item \emph{Traffic Optimization} Traffic optimization is a broad topic covered from the optimum traverse route planning to driver assistant system. And automatic collision avoidance assisted by VANETs is also under this theme.
\item \emph{Network Security} Network security has been a long standing problem since the creation of the internet. In VANETs, there are obviously different features for this problem compared with that in internet or wireless networks in general sense. IPT can be effectively applied for solving this.

\end{itemize}

IPT can be applied into VANETs to support applications mentioned above. It should be noticed that IPT's type are many. To classify different kind of IPT and overview their applicants in VANETs separately dose make sense. In \ref{sec ipt classification} we try to investigate the application status of different type of IPT to different type of VANETs applications, mainly from a technical point of view. How to industrialize IPT in VANETs is another issue of concern. \ref{sec relaization road-map} gives a detailed description.

Abbreviations used throughout the paper are described in Table \ref{tab1}.
\begin{table}[htbp]
\centering
\caption{Abbreviations used throughout the paper}\label{tab1}
\begin{tabular}{|l|p{6.5cm}|}

\hline
ACO &Ant Colony Optimization Algorithm \\
\hline
ADAS &Advanced Driver Assistance Systems\\
\hline
BTSC &Bus Trajectory Based Street Centric \\
\hline
CW &Contention Window\\
\hline
COG &Center Of Gravity \\
\hline
CART &Classification And Regression Tree\\
\hline
CHAID &CHi-squared Automatic Interaction Detection\\
\hline
GPS &Global Position System\\
\hline
IEEE &Institute of Electrical and Electronics Engineers\\
\hline
IPT &Intelligent Processing Technology\\
\hline
ISO &International Organization for Standardization\\
\hline
ML &Machine Learning\\
\hline
OBU &On Board Unit\\
\hline
P2P &Peer-to-Peer\\
\hline
QoS &Quality of Service \\
\hline
RSS &Received Signal Strength \\
\hline
RSU &Road-Side Unit \\
\hline
SDO &Standards Development Organization \\
\hline
V2I &Vehicle-to-Infrastructure \\
\hline
V2V &Vehicle-to-Vehicle \\
\hline
V2R &Vehicle-to-Road \\
\hline
R2R &Road Side Unit-to-Road Side Unit \\
\hline
VANET &Vehicular Ad-hoc NETwork \\
\hline
\end{tabular}
\end{table}

\subsection{IPT Classification based on Application in VANETs}
\label{sec ipt classification}

\begin{figure*}[htbp]
\centering
\includegraphics[width=1\textwidth,]{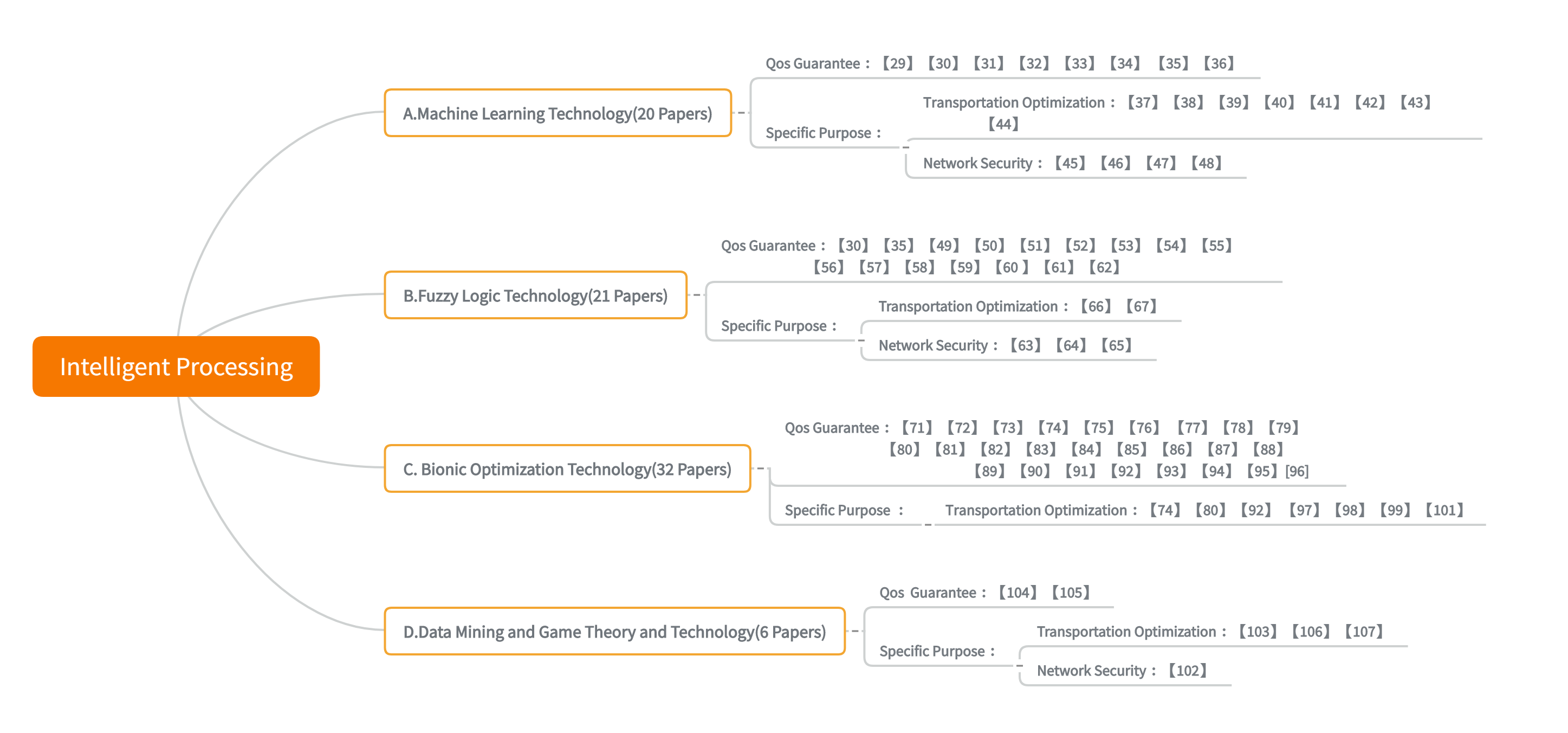}
\caption{Classification of intelligent processing technologies}
\label{figIP}
\end{figure*}

In Fig\ref{figIP} we classify IPT cited by this paper based on their categories and targets in VANETs.

What can be clearly seen in this figure is the high rate of machine learning technology and fuzzy logic technology. The former is mainly used to guarantee QoS, avoid traffic congestion and collision. Machine learning technology can provide systems the ability to automatically learn and improve from experience without being explicitly programmed. That is to say, vehicles have predict power to some extent. Fuzzy logic technology, which is simple and easy to a realization, imitate human decision process and has very practical implications.

\begin{figure}[htbp]
\centering
\includegraphics[width=8.5cm]{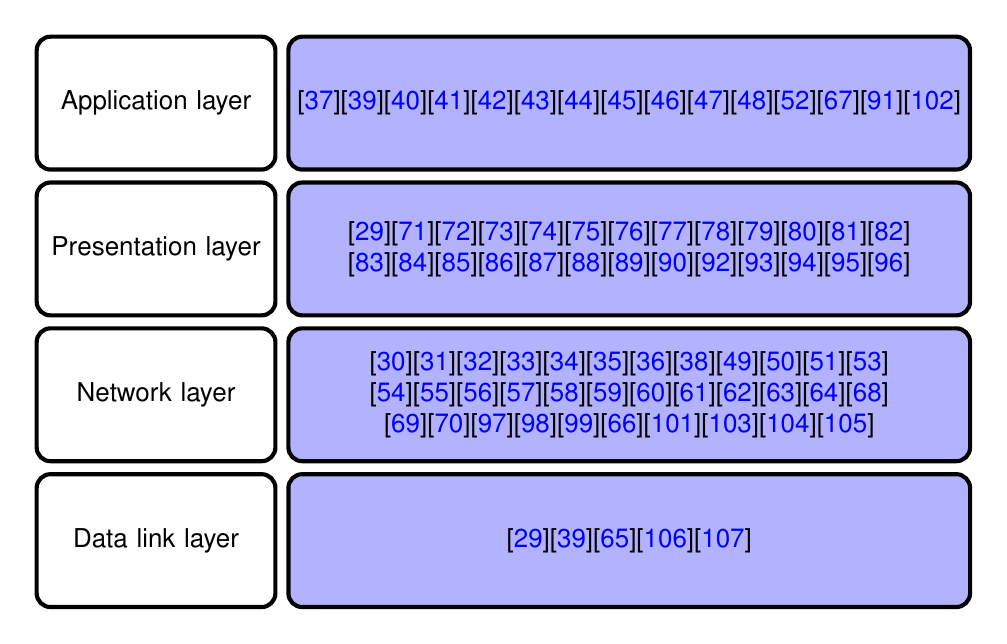}
\caption{IPT Usage in VANETs from the Perspective of OSI Reference Model}
\label{figOSI}
\end{figure}

From another perspective, based on the network protocol layer the VANETs application belongs to, we can classify the cited works as shown in Figure \ref{figOSI}. What stands out in Figure \ref{figOSI} is the concentration on application,transport,and network layer.  The causes of this phenomena can be explained in two ways. Firstly existed techniques in physical and data link layer are mature and reliable, just adapt them can reduce the difficulty of realization. Secondly, in physical and data link layer, there generally does not exist multivariable complex decision-making, so IPT is seldom used.

\subsection{IPT Realization Road-Map}
\label{sec relaization road-map}
The industrialization of new IPT is a lasting and complicated process that may take years and involve many international organizations and professional associations and government regulators. It is meaningful to give the concrete steps of this process clearly. Generally speaking, the procedure is similar to the situation in traditional wireless networks. But it is important to note that, for now, there is no fixed international standard for VANETs. The whole process should be tracked, adjusted and monitored by a Standards Development Organization (SDO), such as IEEE, IEC, ISO, and others. Here we take the policy of IEEE as an example.

\begin{figure*}[!tbp]
\centering
\includegraphics[width=1\textwidth]{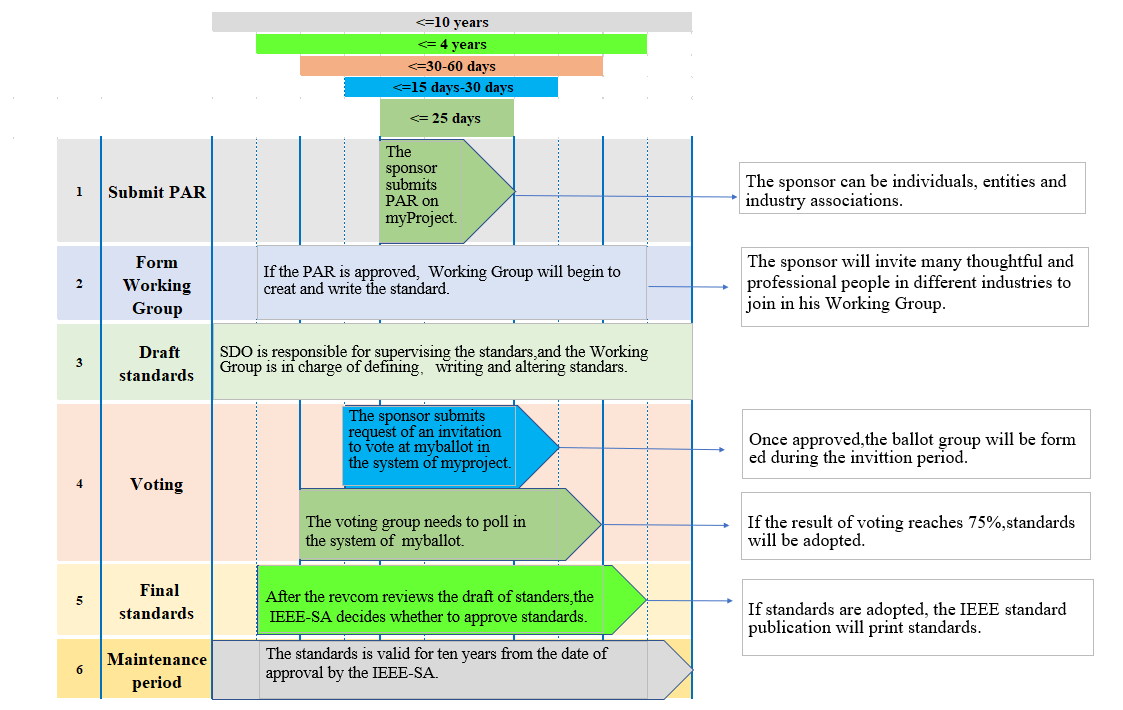}
\vspace{-0.2cm}
\caption{The Roadmap for Industrializing IPT}
\label{figT}
\end{figure*}%

What stand out in Figure \ref{figT} is the general organization or department responsible and the processing time needed in each step. They are listed as follows:

\begin{itemize}
\item \emph{Step 1 Submission of request PAR (25 days)}

\item \emph{Step 2 Set Up Working Group}


\item \emph{Step 3 Preparation of the Draft}


\item \emph{Step 4 Casting Ballot (invitation 15-30, voting 30-60 days)}


\item \emph{Step 5 Obtain final criteria (Less than or equal to 4 years)}：


\item \emph{Step 6. Revised, corrected and maintained Standards (10 years)}
\end{itemize}

\section{Machine Learning Technology in VANETs}
\label{sec:NeMl}
Machine learning (ML) is a general term for the combination of algorithms and statistical models focusing on specific task that a computer system facing.

ML can be roughly divided into these categories:
\begin{itemize}
  \item Regression. Include Ordinary Least Square, Logistic Regression, Stepwise Regression,Multivariate Adaptive Regression Splines, and Locally Estimated Scatterplot Smoothing.
  \item Decision tree. Include Classification And Regression Tree(CART),Iterative Dichotomiser , Chi-squared Automatic Interaction Detection(CHAID),Decision Stump,Random Forest,Random Forest,and Gradient Boosting Machine.
  \item Bayes Methods. Include Naive Bayes, A veraged One-Dependence Estimators Classification, and Bayesian Belief Network.
  \item Kernal Based Methods. Include Support Vector machines and Linear Discriminant Analysis.
  \item Clustering Methods. For example, K-Means.
  \item Artificial Neural Networks. Include Perceptron Neural Network, Back Propagation, Hopfield Network, Self-Organizing Map, and Learning Vector Quantization\cite{Zhang2018}.
  \item Deep Leaning. Include Restricted Boltzmann Machine, Deep Belief Networks, Convolutional Network, and Stacked Auto-encoders.
  \item Dimensional reduction. For example, PCA, MDS, Manifold Learning, and so on\cite{Chen2011a,Chen2012a,Chen2013b}.
\end{itemize}

In essence, ML provide a series of utilities to  find out patterns and features efficiently from large numbers of data. These patterns and features can inform decision-making in different layers in Ad Hoc Networks and VANETs\cite{Chen2013c,Liu2013a}.

\subsection{ML for QoS Guarantee}
QoS means Quality of Service, which is the description of overall performance of the network, or the ability to guarantee a certain level of performance to a data flow. Key indicators include network throughput, end-to-end delay and packet deliver rate. A lot of ML methods used in VANETs can be grouped together because their aims are QoS guarantee. For example, all routing protocol design effort can be considered as QoS guarantee goal orientation.

In VANETs QoS oriented application, machine learning technology always play a role in decision making when facing a relative long decision period and there is enough data. From the data some pattern will be recognized and optimum decision will be made.

In this topic, we can break the ML application in VANETs further into MAC layer\cite{ISI:000406600300093} and Network Layer\cite{AnWuYoshinagaEtAl2018,ISI:000353155700001,ISI:000433209500011,ISI:000364988900012,ISI:000387902500024,WuYoshinagaChenEtAl2018,ISI:000447510600001}.

\cite{ISI:000406600300093} propose a Q-Learning based back-off algorithm to overcome the problems of  traditional IEEE 802.11p MAC protocol. These problems include low packet delivery rate, high delay and poor scalability in VANETs. In this protocol, the vehicle nodes interact with surroundings continuously and learn from each other. The the length of contention window can be adjusted according to the learning result. With the optimal CW length, the possibility of packet collisions and end-to-end delay are minimized. Q-learning method does not require priori knowledge of the environment so that it is effective. The nodes adjust its behavior without the model of environment. Through maximizing the cumulative reward corresponding to perform special action in a stage, the problem of collision and delay are solved by set optimal CW.

\cite{AnWuYoshinagaEtAl2018} propose a context-aware edge-based packet forwarding scheme for vehicular networks. The last two hop communications use a reinforcement learning algorithm for optimization. Because edge nodes are used to forward packets, the number of hops may increase unnecessary. This negative situation may occur when source/destination node is very close from the edge. In this paper, a Q-learning method is used for each node to select next hop. In the Q-learning model, the action of selecting an one-hop neighbor as the next hop is assigned with a rewarding value, corresponding with the link status. Through special design, the Q-value represents the evaluation of a next packet forwarder candidate with the consideration of multi-hop performance.

\cite{ISI:000353155700001} point out that traditional research works tends to select path in higher vehicle density area in order to avoid carry-and-forward problem, and decrease transmission delay. This strategy may not get accurate status about the traffic since the topology of VANETs is changing rapidly. This paper assume each vehicle equip GPS device and so that all RSUs have the road and location information. In the most complicated scenario, the destination vehicle , is not in the coverage area of any RSU, then a machine learning algorithm is used to predict the location of the destination. Another two prediction should be made are the evaluation of transmission capacity and forwarding direction. The inputs to the predict for the first target are the moving path of the destination of its coverage, the driving lane and spot speed of the destination when it left. For second mission, to select the path to maximize the network capability based on a unsupervised learning method. To calculate the forwarding direction, the machine learning method uses vehicle speeds and transmission capacities of each path, so that wireless links also can be accepted.

\cite{ISI:000433209500011} propose a clustering-based reliable routing algorithm which use radial basis function neural network to select cluster head considering velocity and free buffer size parameters. It use RBF neural network to select cluster head node. After clustering is done, the best node will be selected through the training which was previously given to RBF network. The cost function consider vehicular velocity and vehicular maximum velocity, and free space and initial value for vehicle buffer,respectively. For each cluster member node, the aforementioned cost function is calculated, the obtained value are regarded as the input of the RBF function for each cluster to select the best cluster head with the highest value.

\cite{ISI:000364988900012} applies data classification in order to predict which nodes are the most suitable as intermediate according to temporal and local connectivity. This work train a classifier and get a decision tree and send the tree to each node. The specific node are thus predicted, according to the region where the node will transmit.

\cite{ISI:000387902500024} propose a centralized and localized data congestion control strategy  using RSUs at intersections. Based on size, validity and type, messages are clustered using K-means algorithm, one of the most
popular unsupervised learning algorithms. Then RSUs give feedback information to the vehicles stopped before the red traffic lights to reduce communication collisions.

\cite{WuYoshinagaChenEtAl2018} target on distributing a large amount of content to vehicles on the road. This work propose a two-level clustering approach. In the second level, Q-learning algorithm is used to tune the number of gateway nodes. For Q-learning algorithm, the problem is whether an edge cluster head should work as a gateway or not. The whole network is the environment and the edge cluster heads are learning agents. Each node will select the next hop for data transmission, so the set of neighboring nodes form the possible actions allowed. Each node maintains a Q-Table where each Q-value reflect the node density. The mechanism of update Q-Table reward node balance the next-hop selection between direction connection with BS or neighboring gateway with large number of devices.

\cite{ISI:000447510600001} propose an improved AODV routing protocol based on fuzzy neural network. This neural network is used to calculate the node stability. The metrics of node stability in this work are adjacent node's relative velocity, relative distance, and node load.

\subsection{ML for Specific Purpose}
There are two main categories of "Specific Purpose", Transportation Optimization\cite{Chen2018,WuYoshinagaJiEtAl2017,ISI:000441814300154,ISI:000447216400001,ISI:000413518200001,ISI:000375206200025,ISI:000337130600014,ISI:000373098400004}, and Security\cite{ISI:000361897200019,ISI:000380402400162,ISI:000370305000004,ISI:000382480800004}. Referring to transportation optimization, the topics include but not limited to collision prediction, data storage, vehicle diagnosis, infrastructure establishment etc. Security here does not mean transformation safety, but targeting at security against malicious attacks.

\cite{Chen2018} use deep learning schemes to improve driving safety. The rear-end's decision of serving the chauffeur is determined by BP neural network through evaluating the possible collision risk, whose information is get by VANETs communications.\cite{ISI:000447216400001}  use image sensor to capture and analyze vehicle's driving situation and interact with them. This work track vehicles based on human attention mechanism for self-selection of deep features based on a modified convolutional neural network.

\cite{WuYoshinagaJiEtAl2017} concerning the data storage in VANETs issue. This work try to transfer the interested data to a new vehicle when the current carrier is going to leave the specific region. Fuzzy logic and reinforcement learning technique is used to select the next carrier considering several parameters such as throughput, velocity, and bandwidth efficiency.

\cite{ISI:000441814300154} proposes an integrated self-diagnosis system for autonomous vehicle based on deep learning. A module in this system creates the training dataset on the basis of the data collected by in-vehicle sensors. This module also use deep learning to diagnoses the condition of parts and of the other parts influenced by other parts. The input is the sensor data, the output represents the condition of each part. Based on the part's condition ,a lightweight neural network is used furthermore to diagnoses the vehicle's total condition.

\section{fuzzy logic technology}
\label{sec:flf}

Fuzzy logic is an "lightweight" decision making mechanism which based on "degrees of truth" rather than the usual "true or false" boolean logic. It is rule-based so that can rely on the practical experience of an operator, particularly useful to capture experienced operator knowledge. This technique is well suited to nonlinear and multiple inputs multiple outputs systems. When conventional means to model a problem is hard, fuzzy logic may be a proper alternate solution.


The core concept of fuzzy logic is to change the mind set of binary logic's two-valued result (either 1 or 0) to a continuum of values between 0 and 1. That means to use a percentage to describe a state. There are two important concepts here: membership function and rule. A membership function in fuzzy logic is an extent description of a physical variable. A rule stipulate what output should be give based input given.

For a fuzzy logic fashion decision process, the first step is to define a set of membership functions related to each input and each output. Secondly, these membership functions use predefined rules to yield output value. From graphical aspect, every input's corresponding membership function solution will cut an area in output's membership function graph. One can imagine that multiple inputs will produce an union graph area in multiple output's graph area. The final decision is made by calculating the centroid of this union area. By adjusting the membership function and the rules, the optimum fuzzy system is achieved.

Compared with mathematical optimization methods, fuzzy logic has  many merits such as  low-complexity implementation. This is important to satisfy the quickly responding requirement of the VANETs.

\subsection{Fuzzy Logic for QoS Guarantee}
Using fuzzy logic for QoS guarantee in VANETs \cite{AnWuYoshinagaEtAl2018,ZhiouaTabbaneLabiodEtAl2015,HassanAhmedRahman2013,LogeshwariLakshmanan2017,SoleymaniAbdullahAnisiEtAl2017,WuYoshinagaChenEtAl2018,WuYoshinagaJi2017,MahajanVijayakumar2017,FatemidokhtRafsanjani2018,JadhavDongreDevurkar2017,ISI:000372744600007,AbadaMassaqBoulouz2017,ISI:000361632400167,DhimanJadhav2015,MoridiBarati2017,HuWuZhaoEtAl2018} is a hot topic in recent years.

\cite{AnWuYoshinagaEtAl2018} use fuzzy logic to select edge node as a forward relay to ease off payload of wireless transmission so that achieving efficient use of wireless resources. In the process of edge node selection, vehicle velocity, vehicle distribution, and channel conditions between edge and vehicle are taken into account. The membership function of these factors and output, together with IF/THEN rules are defined, then this work use the Min-Max method to combine all the rules together. Through calculating the center of gravity(COG), the numerical value of competency of being an edge node.

\cite{ZhiouaTabbaneLabiodEtAl2015} use received signal strength, load of the cluster head, and gateway candidates, and vehicle-to-vehicle link connectivity duration as fuzzy logic's criteria to select the gateway for V2I communication. That means, the gateway between the source vehicle and the LTE advanced infrastructure. The source vehicle perform the whole decision process. At data collection phase, the network card of 802.11p and LTE interface , the source vehicle measure link connectivity strength,load of the cluster head and  RSS of neighbouring BSs respectively. In decision Phase, the type of data traffic should be classified, because voice, steaming and data has different tolerance of delay. The fuzzifier transforms the input values into degrees of matching with linguistic values.Input parameters, such as source to infrastructure link connectivity strength, cluster head load,.etc, are fuzzified using the predefined input membership functions. And a set of IF-THEN rules are generated to obtain the fuzzy gateway decision. In the final step, centroid is calculated for defuzzification.

\cite{HassanAhmedRahman2013} propose an adaptive beaconing approach so that vehicles can regulate their beacon rate based on traffic condition. packet carried time, number of single-hop neighbors, and vehicles speed are criteria input for fuzzy logic system. Beaconing interval is the output. Membership function and fuzzy rules are defined and centroid for output is calculated. \cite{SoleymaniAbdullahAnisiEtAl2017} also deal with the beacon rate adaption problem. for this work's fuzzy logic model,traffic density, vehicle status and location status are input criteria. Traffic density is classified into low, medium,and high. And location's status is divided into two situation, hazard and non-hazard. Vehicle's status is divided into emergency and non-emergency. Using fuzzy logic, rate of beaconing is adjusted based on such criteria. This mechanism incorporates essential humanistic concepts.

\cite{WuYoshinagaChenEtAl2018} counts competency of nodes based on their velocity, leadership and signal quality factor. The competency value is use to select cluster head.

\cite{WuYoshinagaJi2017} \cite{HuWuZhaoEtAl2018}selects efficient gateway nodes using fuzzy logic based on velocity, leadership, and antenna height factors, that can insure the . The gateway nodes bridge the licensed Sub-6 GHz communcation and mmWave communication in order to maximize the overall network throughput. Instead of each vehicle connecting to a BS, only the gateway vehicles utilize Sub-6 GHz interface and communicate with other vehicles through mmWave V2V communications.

\cite{MahajanVijayakumar2017} point out the lack of specific optimization on emergency information delivery in current research. A fuzzy logic based scheme is proposed to predict the movement of vehicles, so that reduce the delay of emergency information delivery. All three communication types ,V2V,V2R(Vehicle to Road Side Units), and R2R( Road Side Unit to Road Side Unit) are all used . Because fuzzy logic is robust in nature and is able to work with information which has less precision, noise and uncertainty, it is used to predict vehicle movement. From this prediction, whether accident spot is getting congested or not will be known so that message could be transmitted to the coming vehicle to change their route in advance. In this work, RSU determine vehicle's speed, distance and direction , form corresponding membership functions and rules and select the the appropriate node as the next hop whose speed is low, distance is less and direction is towards the accident spot.

\cite{FatemidokhtRafsanjani2018} propose a route protocol use fuzzy logic to determine trust value which indicating each link's validity. The input criteria is bandwidth, connectivity level and congestion level of the link. Here bandwidth means the number of bytes sent by the nodes. Connectivity level means received signal strength metric which is calculated use physical layer parameters such as antenna gain, transmission range and the wavelength.

\cite{JadhavDongreDevurkar2017}use fuzzy logic to select best forwarding nodes. Input criteria are distance, mobility,RSSI. \cite{AbadaMassaqBoulouz2017} use fuzzy logic to select best relay so that delay and cost can be reduced. The criteria include link stability (mobility) and received signal quality and moving direction.

\cite{ISI:000361632400167}use fuzzy logic to estimate the link quality based on it's delay, packet collision and bandwidth. Modified routing protocol so that can chose best links to improve packet delivery ratio,reduce end-to-end delay and cost. \cite{DhimanJadhav2015} use reliability and survive time of link as fuzzy input to determine link's competence value.

\cite{MoridiBarati2017} use fuzzy logic to extend the AODV routing protocol to create reliable routing between cluster heads. Traditional AODV always try to select route with the minimum hop count. In VANETs, due to high mobility, this choice may be unreliable due to link's broken. This work uses link expiration time(LET) and link reliability. Link expiration time is calculated based on velocity and direction parameters and the angle between two vehicles. Link reliability is information transferability with a minimum link failure as a conditional probability. Unlike other works, this work use a COA method to defuzzification instead of calculating the centroid of output.

\begin{table*}[tp]\normalsize
 \caption{ Comparison of Fuzzy Logic Technology in VANETs}
\resizebox{\textwidth}{30mm}{
 \begin{tabular}{lclll}

  \toprule
 Paper & Strategy & Architecture & Challenge & Application \\
  \midrule
\cite{AnWuYoshinagaEtAl2018} & { Fuzzy Logic }& V2V & Improve the efficiency of the end-to-end communication & Achieve an efficient use of wireless resources \\
\cite{WuYoshinagaChenEtAl2018} & { Fuzzy Logic and Q learning } & V2V & Cellular network is not sufficient due to its limited bandwidth & Improve throughput in high-density scenarios\\
\cite{ZhiouaTabbaneLabiodEtAl2015} & Fuzzy Logic & V2I & Selecting gateway candidate & Better performance in delay and packet loss \\
\cite{SoleymaniAbdullahAnisiEtAl2017} & Fuzzy Logic & V2V & Minimizing control plane modifications & Reduce overhead \\
\cite{SoleymaniAbdullahAnisiEtAl2017} & Fuzzy Logic& V2V,V2I &Emergency messages face a poor performance& Reduces the congestion and increases the information accuracy\\
\cite{FatemidokhtRafsanjani2018} & Fuzzy Logic &V2V&Security threats &Guarantee road safety service quality\\
\cite{ISI:000372744600007} & Data Mining & V2V,V2I &Fair channel allocation schemes among vehicles & Improve transmission reliability and security \\
\cite{AbadaMassaqBoulouz2017} & Fuzzy Logic & V2V,V2I & Improve IEEE802.11p & Improve the routing performances in the network \\
\cite{ISI:000361632400167} & Q-Learning & V2V & Multi hop communication in VANETs & Good performance in packet delivery ratio, end to end delay and overhead \\
\cite{DhimanJadhav2015} & The mathematical model of normal distribution & V2V,V2I & The occurrence of link breakage & Increasing the throughput \\
\cite{MoridiBarati2017} & Fuzzy Logic & V2V & Developing a trap in the local optimum & Reduce link failure and packets loss \\
\cite{HuWuZhaoEtAl2018} & Fuzzy Logic & V2V & Limited bandwidth in a dense vehicle environment & Achieve High overall network performance \\
\cite{BojnordBojnord2017} & Sybil Attack Detection & V2V & Security threats & Propose different novel detection schemes  \\
\cite{SoleymaniAbdullahZareeiEtAl2017} & Fuzzy Logic & V2V,V2I & Interruption caused by information collected by vehicles and obstacles & Detect malicious attackers and faulty nodes \\
\cite{MejriBen-Othman2017} & Fuzzy Logic & V2V & Vulnerable to many DoS attacks especially the greedy behavior & Can be executed by any node and without modification of the IEEE 802.11p standard\\
\cite{ISI:000321700400031} & Fuzzy Logic & V2V & Serious channel competition & Achieve a high accuracy of road congestion detection \\
\cite{NassarKarray2016} & Fuzzy Logic & V2V & Reduce the number of fuzzy sets & Improve  VANET safety services \\

  \bottomrule
 \label{tab:Date}
 \end{tabular}}
\end{table*}

\subsection{Fuzzy Logic for Network Security}
\cite{BojnordBojnord2017} focus on Sybil attack in VANETs, use fuzzy logic to form a robust detection mechanism. This work use the concept named node opinion and network opinions. A note's malicious level is computed by the previous node and network opinion. Network opinion is updated based on the previous value of network opinion and sum of opinion of neighbor nodes. Each node in a periodic time runs fuzzy logic to determine malicious level of neighbor nodes.

\cite{SoleymaniAbdullahZareeiEtAl2017} propose a trust model using fog computing that not only detect malicious attackers and faulty nodes, but also tackles the uncertainty and imprecision of data in the vehiculare network in both LOS and NLOS states. Each vehicle individually measures the trust level of the sender of an event message by performing fuzzy logic . For security, VANET needs to fulfil the requirements include authentication, message integrity, confidentiality,location validation and availability. Here fuzzy logic is used to access the accuracy and integrity of a sender of the event message. Once received an event message, the node must evaluate whether the sender is authorized or not, and the event message's lifetime. The accuracy level of the location of the event included in the message is checked by fuzzy logic. Then fuzzy logic is used to evaluate the trust value based on experience, plausibility and accuracy level of location. When sender advice evaluated as reliable, the experience of the history interaction between the sender and current node are calculated. Plausibility level is calculated through location verification of the sender. Accuracy level is measured use fog node. Finally the trust level is determined using fuzzy logic with the experience level, plausibility level and accuracy level.

\cite{MejriBen-Othman2017} propose a new detection algorithm to respond greedy behavior attack. In this algorithm's decision stage, fuzzy logic mechanism is used to determine if a node is either greedy or it is honest to decrease the rates of both false positive and false negative. The greedy behavior in a VANET in this work is defined that nodes do not respect MAC layer access method requirements by manipulating several parameters. By using these manipulations, an attacker penalize the other nodes.The duration between two successive transmissions, transmission time ,connection attempts number of a node are monitored. A greedy node 's duration between two successive transmissions is almost close to zero and occupies the medium more than other normal nodes. The greedy node also tries much more than other nodes to connect to the network. Considering these characteristics, in algorithm's decision phase, the fuzzy input membership function are the number of connection attempts, the average of connection duration, the average of waiting times between connections. Finally, this work also use the most used defuzzification techniques exist, center of thew gravity.
\subsection{Fuzzy Logic for Transportation Optimization}
\cite{ISI:000321700400031} use fuzzy logic for information fusion used in road congestion detection in VANETs. To be specific, a new multilevel information fusion approach including fuzzy clustering-based message aggregation scheme is proposed. The atomic messages are classified into different message clusters by measuring the differences between the atomic messages according to their fuzzy similarity relations. Henceforth,the atomic message in the same message cluster can be aggregated into one,which is called feature information. Obtained message clusters correspond to different traffic features,which reflect the real road conditions. In the fuzzy clustering method, there is a step of creating the fuzzy similarity relationship matrix based on the selected mathematical method. The fuzzy similarity degree between two different message objects is computed using classic arithmetic average minimum method.

\cite{NassarKarray2016} use fuzzy logic to perform crash notification. Vehicle density and vehicle speed are used as input membership function and the output is the estimation of the crash severity.
A comparison of several methods is shown in Table \ref{tab:Date}.


\section{Bionic Optimization, Data Mining and Game Theory Technologies}
\label{sec:Bot}

\subsection{Bionic Optimization Technology}
Bionic optimization technology simulates the evolutionary synthesis. We classify it as one of the intelligent processing technology. The algorithm has the tendency to converge to the best solution if there are enough parents,kids,and generations,mutation radius,even if there are many local optima.

\begin{figure}[h]
\centering
\includegraphics[width=0.3\textwidth]{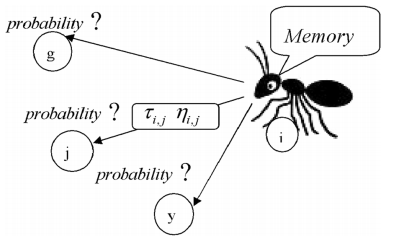}
\caption{The selection of the node\cite{ISI:000375693600004}}
\label{figACO}
\end{figure}

One category of Bionic Optimization, Ant Colony Optimization technology is applied widely in VANETs because of it's good property. An ant colony consists of the ants with relatively simple responses to environment without global control. The ants organize themselves by direct communication. Simple self-organization can accomplish complex tasks. The basic concept of ACO is to divide the whole decision process as multiple iterations. At the beginning, individuals choose one action randomly and complete it, with the knowledge of the action choice's property. Then the knowledge is shared among the colony, so each action choice's benefits are updated and in next iteration, individuals will tend to choose the action with highest benefit, so that the final group optimal result will be achieved\cite{ACO,Gui2016,Gui2016a}. Figure \ref{figACO} shows shows an ant arriving in node i. In VANETs, ACO is mainly used for QoS routing\cite{AzzaliGhazaliOmar2017,Abbas2018,Correia2011,Davoodi2015,Eiza2016,Eiza2015,Fatemidokht2018,Ghorai2017,Hegde2016,Jindal2018,Jindal2015,Jyothi2018,Kaur2016a,Kazemi2013,Khoza2018,Li2014,Li2013,Li2017,Majumdar2016,Mane2013,Sun2018a,Thilak2018,Wahab2013,Zhang2017c,Zhang2017a,Zhang2017b}
, another usage is for transportation optimization \cite{Kponyo2014,Davoodi2015,Jindal2016a,Thilak2018,Jindal2018,Goudarzi2019}

\cite{Goudarzi2019} presents a traffic-aware position-based routing protocol for VANETs. This work use ACO to count the sum weight of the whole route.\cite{Sun2018a} propose a bus-trajectory-based street-centric(BTSC) routing algorithm, uses buses as the main relay to deliver messages. The route selecting strategy consider higher density of busses and lower probability of transmission direction deviating from the routing path.\cite{Abbas2018} employ ACO to efficiently compute the optimal routes considering reliability, end-to-end latency, throughput and energy consumption.\cite{Jyothi2018} modify traditional ACO and dynamically makes decision in choosing shortest best route in highly congested areas.\cite{Khoza2018} increase the efficiency and reliability of vehicle traffic information message transmission by an Ant Colony Hybrid Routing Protocol, which result in better packet delivery ratio and end-to-end delay.\cite{Zhang2017a} points out that ACO is useful for improving reliability of VANETs by obtaining several alternatives.\cite{Li2017} use ACO to compute the route selection problem, which is formulated as a constrained optimization problem.

\cite{ISI:000426011800037} proposed an improved hybrid ant particle optimization algorithm to optimize the route to reduce the travel time. When there is an congestion, the route selection mechanism can avoid the hot point and recover when the congestion disappeared.

\subsection{Data Mining and Game Theory Technology}

Data technology search for hidden information in large amount of data. Game theory technology imitate player's strategy in a game to get the optical reward. These two technologies are proven intelligent processing technologies, but relatively seldom used in VANETs for now. Rather than  exhaustive, we just list several typical applications below.

\cite{ISI:000427342300002} proposed a game theory based intrusion detection framework. The interaction between the IDS and the malicious vehicle is modeled as an two player non-cooperative game and the IDS monitoring strategy is based on the Nash Equilibrium of the game. This model makes the volume of IDS traffic minimized.

\cite{Wu2018} uses game theory based reward allocation mechanism for a reinforcement learning algorithm to route selecting of each vehicles.\cite{Wang2018} design a inter-vehicle cross-layer cooperative game model taking into account the global optimal utility of the game players. The aim of this mechanism is to improve the application value defined in the value of a conveyed packet.\cite{AbuShattal2018} focus on dynamic spectrum access, using evolutionary game to select the strategy for resource contention.

\cite{Ruta2018} address the problem of high-level context informaton sharing,and \cite{Zhang2018a} proposes a framework enabling a contextual data management and mining in VANETs.

\section{Conclusion}
\label{sec:CFD}
Intelligent Processing Technology imitate human thought to some extent. It has relatively low computation complexity , easy for application, and at the same time, maintain reasonable accuracy. For VANETs, because of the rapidly changing topology and strict constraint for QoS, IPT is a reasonable choice for a lot of applications.

As show in this review, IPT include Machine Learning, Fuzzy Logic, Bionic Optimization, Data Mining and Game Theory technologies are widely used in VANETs for QoS , transportation optimization and Security. Different layers of network may be involved and different kind of IPT technologies can be combined to work together. We also give a brief summary of the roadmap about industrializing a specific IPT technology in VANETs. The study contributes to our understanding of VANETs.

The proper mathematical modelling of the used IPT and their precise comparison in different aspects with the mathematical optimization methods are also important. The lack of deep study in this regard may be said that IPT for VANETs study is still superficial. This would be a fruitful area for further work.

\section{Acknowledgment}
This work was supported in part by the Nature Science Foundation of Guangdong Province(2014A030310266).

\bibliographystyle{plain}
\small
\bibliography{references1}
\end{document}